\documentclass[twocolumn,prl,aps,showpacs]{revtex4}

\usepackage{graphicx}
\usepackage{color}
\def\be{\begin{equation}}
\def\ee{\end{equation}}
\def\bea{\begin{eqnarray}}
\def\eea{\end{eqnarray}}
\def\ba{\begin{array}}
\def\ea{\end{array}}
\def\bdm{\begin{displaymath}}
\def\edm{\end{displaymath}}

\begin{document}

\title{Dimerized and trimerized phases for spin-2 Bosons in a one-dimensional optical lattice}

\author{Pochung Chen$^1$, Zhi-Long Xue$^1$, I. P. McCulloch$^2$, Ming-Chiang Chung$^3$, and S.-K. Yip$^4$}
\affiliation{$^1$Physics Department, National Tsing Hua University, Hsinchu, 30013, Taiwan}
\affiliation{$^2$School of Physical Sciences, The University of Queensland, Brisbane, QLD 4072, Australia}
\affiliation{$^3$Physics Division, National Center for Theoretical Sciences, Hsinchu, 30013, Taiwan}
\affiliation{$^4$Institute of Physics, Academia Sinica, Taipei 11529, Taiwan}
\date{\today }

\begin{abstract}

We study the phase diagram for spin-2 bosons loaded in a one-dimensional optical lattice.
By using non-Abelian density matrix renormalization group (DMRG) method we identify
three possible phases: ferromagnetic, dimerized, and trimerized phases.
We sketch the phase boundaries based on DMRG. We illustrate two methods for identifying
the phases. The first method is based on the spin-spin correlation function while
in the second method one observes the excitation gap as a
dimerization or a trimerization superlattice is imposed.
The advantage of the second method is that it can also be easily implemented in
experiments. By using the scattering lengths in the literature we estimate that
$^{83}$Rb, $^{23}$Na, and $^{87}$Rb be ferromagnetic, dimerized, and trimerized respectively.

\end{abstract}

\pacs{67.85.Fg,03.75.Mn,67.85.Hj}





\maketitle

Cold atomic gases have been actively studied in recent years for they offer us
new possibilities of studying quantum many-body systems \cite{Bloch08}.
In the very early experiments of dilute Bose gas in a trap,
Bose-Einstein condensation was beautifully observed directly \cite{BEC}.
We have also witnessed the realization of the Bose-Hubbard model and the observation
of the superfluid-Mott transition \cite{Greiner02},
a phenomenon theoretically predicted long ago but only observed recently.  In this case,
the presence of a lattice and the interatomic interaction actually destroy the superfluid,
resulting in an ``insulating'' state.  In \cite{Greiner02} and also the many
following experiments, the Bosons are spin-polarized and so effectively spinless.
However, Bose-Einstein condensation of bosons with spin degree of freedom 
has also been realized \cite{S1}.  Hence it is natural to ask what would be the
spin ordering of such Bosons in a lattice in the Mott-insulating state, where
even though one is confined to integer number of particles per site, the spins of
the neighboring sites can still interact via virtual tunneling. 
Indeed, this question is of much theoretical interest, as it can be easily
shown that the effective spin-Hamiltonians one can realized
for spinor bosons loaded in an optical lattice are
very different from the Heisenberg-like Hamiltonians much studied in
electronic systems \cite{SH}.
Similarly, a two-component Bose system allows
us to  realize the XXZ spin-1/2 model \cite{XXZ} much discussed in the theoretical literature.


In this paper, we consider spin-2 Bosons in a one-dimensional lattice.
Spin-2 systems are already available and experimentally
studied \cite{Schmaljohann04,Kuwamoto04,Widera06,Tojo09}.
The theoretical phase diagram of spin-2 condensates
is a function of the scattering lengths $a_S$ in the spin
$S=0$, $2$, $4$ channels \cite{Ciobanu00,Koashi00}.
It is divided into three regions, which are named
ferromagnetic (F), polar (P) and cyclic (C) in Ref.~\cite{Ciobanu00}.
For spin-2 Bosons with one particle per site in a higher dimensional lattice
in the insulating phase, it can easily be shown that again the phase diagram is
divided into three regions in the mean-field limit,
in direct analogy to the Bose-condensed case \cite{Zhou06,Barnett06,notes,Yip07}
(see also \cite{fluct}).
In one-dimension (1D), however, strong quantum fluctuations
are expected to substantially modify the phases.  In particular,  the polar and
cyclic phases are no longer expected to be stable.  These states break
rotational symmetry, which implies the existence of linear Goldstone modes,
and thus have diverging quantum fluctuations in 1D.

In this paper, we use non-Abelian density matrix renormalization group (DMRG) method
to determine the general phase diagram for one particle per site.
This regime is much more stable than multi-particles per site and thus
has a much better chance of being realized experimentally.
We find three phases, which are ferromagnetic (F), dimerized (D) and trimerized (T) phases.
The ferromagnetic state
has a macroscopic spin, and has large degeneracies arising from the choice
of the spin projections.
 The dimerized phase has spontaneously broken lattice symmetry, with unit cell consisting of two lattice sites.
 The ground state is a spin singlet,
 with finite gaps to the first excited states.
The trimerized phase is the most intriguing.  At finite size with
total number sites $N$ multiples of $3$, the system is gapped with a spin singlet ground state.
The gap however approaches zero as $N \to \infty$, resulting in a gapless phase.
In this $N \to \infty$ limit, there is also {\it no} broken lattice symmetry.
In the following we shall discuss this general phase diagram and
present numerical evidence leading to our claims.
We further discuss the expected ground states for some available spin-2
elements and how the dimerized and trimerized phases can be obtained experimentally
as well as being tested.

We begin with the Hamiltonian.  Assuming only nearest neighbor interaction,
it is given by $H = \sum_{i=1}^{N-1} H_{i,i+1}$ where $i$'s denote the sites
in increasing order, and $N$ is the total number of sites. 
$H_{i,i+1}$ can be written as
\be
H_{i,i+1} = \epsilon_0 P_{0,i,i+1} + \epsilon_2 P_{2,i,i+1} + \epsilon_4 P_{4,i,i+1},
\label{H}
\ee
where $P_{S,i,i+1}$ denotes projection operators for sites $i$ and $i+1$ onto
a state with total  spin $S$.  Within second order perturbation theory in
the hopping $t$ between nearest neighbors, $\epsilon_S = - 4 t^2/U_S$
where $U_S$ is the Hubbard repulsion for two particles with spin $S$ on
the same site.  $U_S$ is proportional to $a_S$, the s-wave scattering length
in the spin $S$ channel.   For one-particle per site to be stable,
we then need $U_S > 0$ for all $S= 0, 2, 4$, and hence
$\epsilon_0, \epsilon_2, \epsilon_4 < 0$. 
Within DMRG, however, it is more convenient to explicitly express $H_{i,i+1}$
in terms of spin-2 operators $\mathbf{S}_i$, resulting
\begin{equation}
\label{H_spin}
 H_{i,i+1}=\sum_{n=1}^4
  \alpha_n(\epsilon_0, \epsilon_2, \epsilon_4) \left(\mathbf{S}_i \cdot \mathbf{S}_{i+1} \right)^n,
\end{equation}
where
$\alpha_1 = -1/3\epsilon_0 -20/21\epsilon_2 +1/48\epsilon_4$,
$\alpha_2 = -17/180\epsilon_0 -1/9\epsilon_2 +1/40\epsilon_4$,
$\alpha_3 = +1/45\epsilon_0 -1/18\epsilon_2 +1/180\epsilon_4$,
and
$\alpha_4 = +1/180\epsilon_0 -1/126\epsilon_2 +1/2520\epsilon_4$ \cite{noteH,Zang10,quad}.
Since it is crucial to identify the total spin for ground and excited states,
we use non-Abelian DMRG \cite{DMRG} which allows us to find the lowest energy state in different
total spin sectors: $S_{\text{tot}}=0,1,2,3,4,$ and $N$.
In this work we take $N$ up to $60$ with open boundary condition (OBC),
the number of states kept $m=400$ and perform more than $20$ sweeps to ensure the convergence.
We note that though Eq.~(\ref{H_spin}) is more useful for numerical computation,
it is more convenient to discuss the physics using Eq.~(\ref{H}), which we shall do below.

\begin{figure}[tbp]
\begin{center}
\includegraphics[width=0.9\columnwidth]{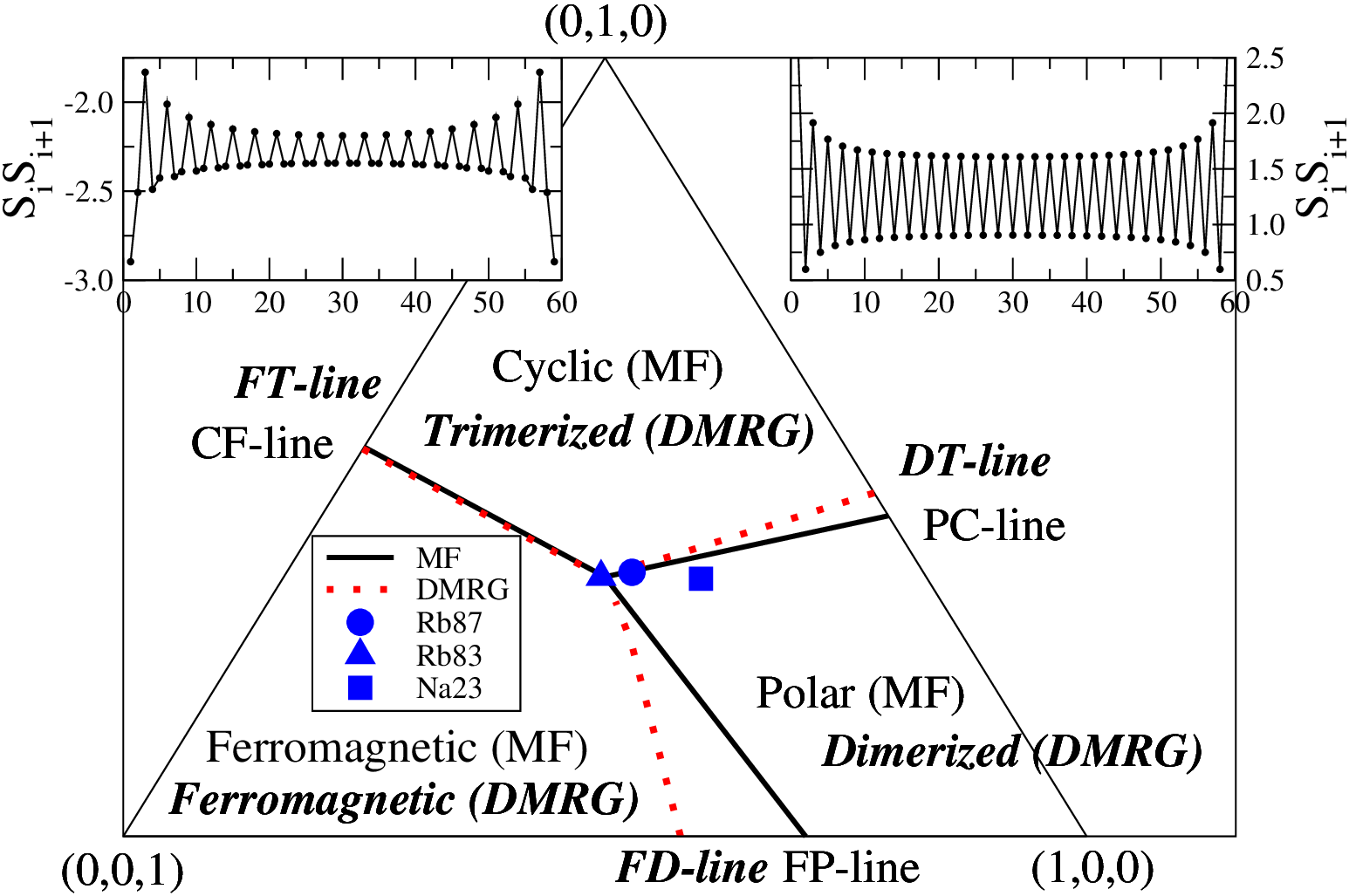}
\caption{(Color online)
Phase boundaries obtained by MF (black solid lines) and DMRG (red dotted lines).
Upper-left: spin-spin correlation for the $(0,1,0)$ point.
Upper-right: spin-spin correlation for $^{23}$Na.  ($N=60$).
The location of $^{23}$Na, $^{83}$Rb, and $^{87}$Rb within this parametrization
are indicated by blue square, triangle, and circle respectively.
}
\label{fig:phases}
\end{center}
\end{figure}

Since the phase diagram cannot depend on the overall energy scale,
we plot the phase diagram in terms of the new variables
$(x_0,x_2,x_4) \equiv (\epsilon_0,\epsilon_2,\epsilon_4)/(\epsilon_0+\epsilon_2+\epsilon_4)$.
By definition $0 \le x_{0,2,4} \le 1$ and $x_0 + x_2 + x_4 = 1$.
It is therefore convenient to present our results using ternary phase diagram,
where $x_{0,2,4}$ are depicted as positions in an equilateral triangle.
We place the points $(0,0,1)$, $(1,0,0)$, $(0,1,0)$ at the lower left, lower right,
and upper corner of the triangle. The coordinates
$(x_0,x_2,x_4)$ at a given point within the triangle are to be read-off by drawing
parallel lines towards the edges and then reading off the intersections.
Our main result is as shown in Fig.~\ref{fig:phases}.

To describe and understand the results, we begin by giving the mean-field phase
diagram as a useful reference.  As mentioned, this can be found in direct analogy
with the Bose-condensed case \cite{Ciobanu00,Koashi00} (see also \cite{Zhou06,Barnett06,fluct}).
The stability region for each phases are (in the form of \cite{Ciobanu00})
\bea
{\rm F}: & & \epsilon_2 - \epsilon_4 > 0, \qquad
  (\epsilon_0 - \epsilon_4) + \frac{10}{7} (\epsilon_2 - \epsilon_4) > 0 \nonumber \\
{\rm P}: & & \epsilon_0 - \epsilon_4 < 0, \qquad
  |\epsilon_0 - \epsilon_4| > \frac{10}{7} |\epsilon_2 - \epsilon_4| \label{mf} \\
{\rm C}: & & \epsilon_2 - \epsilon_4 < 0, \qquad
  (\epsilon_0 - \epsilon_4) - \frac{10}{7} (\epsilon_2 - \epsilon_4) > 0 \nonumber
\eea
These are plotted as black-solid lines in Fig.~\ref{fig:phases}. F, P, and C occupy respectively
the lower left, lower right, and upper part of the triangle.  Since the mean-field energies
of the phases must be linear functionals of $\epsilon_S$'s, the phase boundary are
straight-lines.  They all originate from the point where $x_0=x_2=x_4=1/3$, the center
of the triangle, where all three phases are degenerate.  The intersection points
at the edges are:  FP: $(17/24,0,7/24)$, PC: $(10/17,7/17,0)$, CF: $(0,1/2,1/2)$.


Now we proceed to describe our phase diagram based on DMRG.
Before we discuss each phase in detail,  we summarize our finding as follows:
The polar phase is replaced by the dimerized phase, while the cyclic phase is replaced
by the trimerized phase. The FD and DT lines (no longer straight-lines) are shifted away
from the FP and PC lines but the
FT line remains indistinguishable with the CF line.
The phase diagram is easiest to understand for the lower left region
where $x_4 > x_{0,2}$.  It is
clear that the system would like to acquire the largest possible total spin.  For $N$ sites,
the total spin is $S = 2N$.  For this state, any two neighboring sites have total spin $4$
and the bond energy is $\epsilon_4$, the smallest possible value.  The ground state is
thus $4N+1$ fold degenerate. From our DMRG calculations we find
that the stability region for this state F actually extends slightly beyond $x_4 > x_{0}$.

For the rest of the phase diagram, it is more convenient to first consider
finite $N$.  Later, we shall mention how these pictures are modified as $N \to \infty$.

\begin{figure}[tbp]
\begin{center}
\includegraphics[width=0.9\columnwidth]{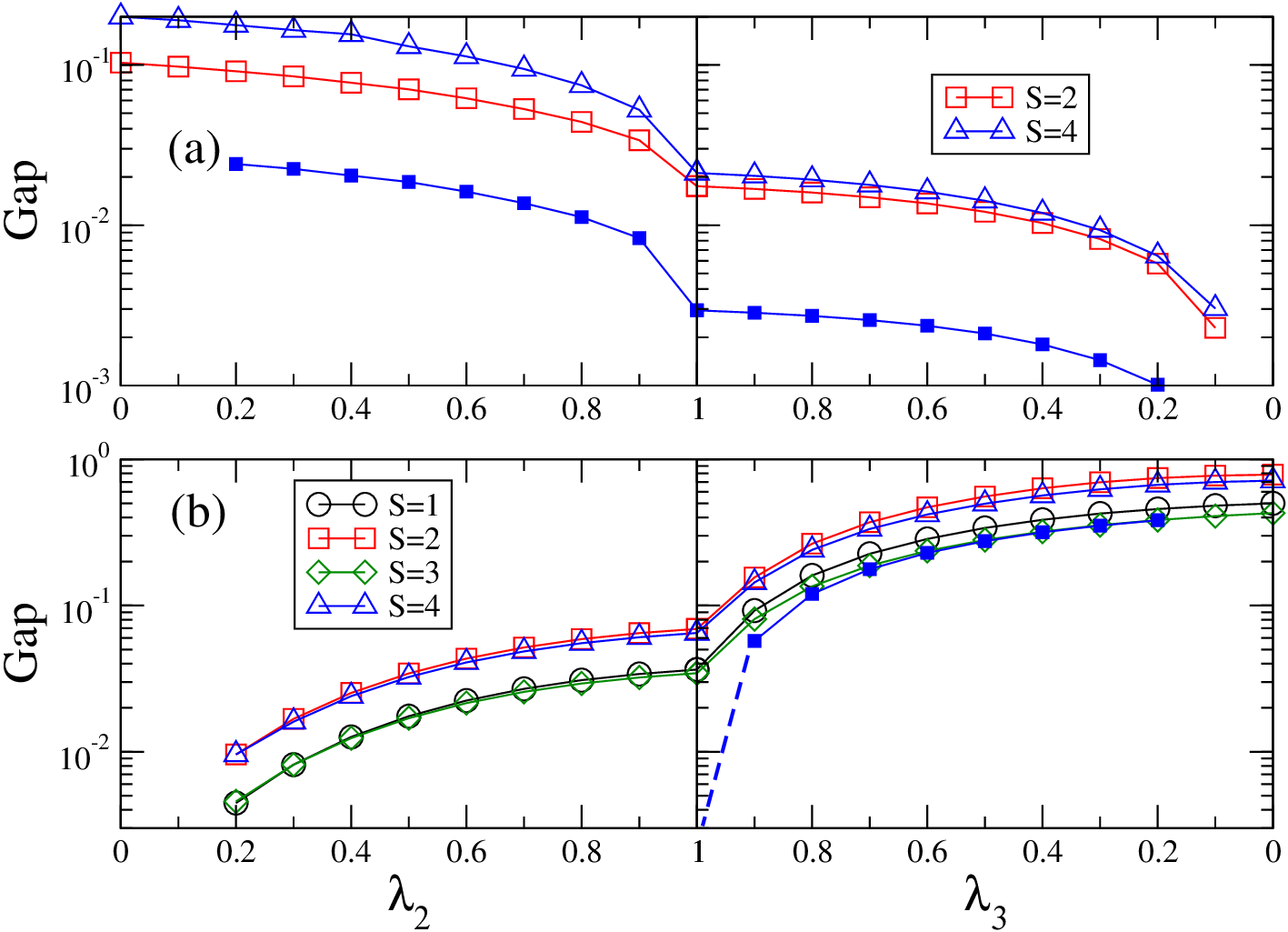}
\caption{(Color online)
Excitation gaps to first few excited states with total spin $S$ for $^{23}$Na (upper plot)
and $(0,1,0)$ point (lower plot) when a dimerization (trimerization)
superlattice of strength $\lambda_2(\lambda_3)$ is imposed. Open symbols:  $N = 60$;
Full symbols: lowest excited state for $N \to \infty$.}
\label{fig:lambda_gap}
\end{center}
\end{figure}

We find that the lower right region of the triangle is occupied by the dimerized phase,
where the unit cell is doubled, and the system is in a non-degenerate singlet state.
In upper-right inset of Fig.~\ref{fig:phases} we provide an example for the spin-spin correlation function
between neighboring sites, where the parameters of $^{23}$Na are used.
The correlation function clearly shows a ``strong-weak'' dimer pattern.
This phase is in direct analogy with the spin-1 case \cite{1Ddimer}.
The ground state for $N=2p$ sites can be most simply understood by first
considering the case of $x_0 \approx 1$ and imposing an artificial dimerization superlattice
where alternate bonds are weakened by a factor
$0 \le \lambda_2 \le 1$ ({\it i.e.}, $H_{2i,2i+1} \to \lambda_2 H_{2i,2i+1}$).
When $\lambda_2 = 0$, the system breaks into $N/2$ subsystems, each consisting of
only two interacting sites. For $x_0 \approx 1$, these two sites form a singlet
with a finite gap to the first excited state(s), and the system is maximally dimerized.
As one increases $\lambda_2$, one expects that the gap should decrease gradually.
In Fig.~\ref{fig:lambda_gap}(a) (left panel) we plot the gap for the first two excited states
as a function of $0 \le \lambda_2 \le 1$. We observe that the gap decreases
monotonically with increasing $\lambda_2$ but never vanishes, indicating the original ground state
at $\lambda_2=1$ is adiabatically connected with that at $\lambda_2=0$ and
hence indeed dimerized.

Next we focus on the upper corner of the triangle, where we find
the trimerized phase.
For $N = 3p$, we find that the ground state is a spin singlet,
with a finite gap to the first excited state.
In upper-left inset of Fig.~\ref{fig:phases} we provide an example for the spin-spin correlation function
between neighboring sites for the case of $x_2=1$. It shows a ``strong-strong-weak''
pattern, indicating the ground state is trimerized.
To get a physical picture of this trimerized phase, it is helpful to consider a three-spins
system in the limit where $x_2 \approx 1$.
For two interacting spins, the ground state is five-fold degenerate, belonging to  $S=2$.
For three spins (say $1$, $2$ and $3$) however, the ground state is a unique singlet:  the total spin of
any two spins can be $0, 1, 2, 3, 4$, and only the spin-2 combination can be added to
the third spin to form a singlet.  By the same argument, this state is an eigenstate
for both the operators $H_{12}$ and $H_{23}$, with eigenvalues $\epsilon_2$.
Hence the system has total energy $2 \epsilon_2$, the lowest possible value.
For system with $N=3p$ sites, it is again helpful to impose a trimerization superlattice
where one every three bonds are weakened by a $\lambda_3$ factor
($H_{3i,3i+1} \to \lambda_3 H_{3i,3i+1}$, $0 \le \lambda_3 \le 1$).
For $\lambda_3 \to 0$, the system break up into $N/3$ subsystems,
each one with a singlet ground state just described and the whole system
is maximally trimerized.
The system at $\lambda_3=0$ is hence gapped, and is stable towards increasing $\lambda_3$ from $0$.
In Fig.~\ref{fig:lambda_gap}b (right panel) we show the gap as a function of $\lambda_3$. We find that
the gap decreases monotonically with $\lambda_3$ but always remains finite, indicating that
the ground state is adiabatically connected to the trimerized ground state of the $\lambda_3=0$ limit.

It is also instructive to study the behavior of the gap when a wrong superlattice is imposed.
Consider a state in the dimerized (trimerized) phase. If one imposes a trimerization (dimerization)
superlattice with strength $\lambda_3(\lambda_2)$, the system will be converted
into trimerized (dimerized) phase as $\lambda_3(\lambda_2) \rightarrow 0$.
As a result one expects that there is
a qualitative change in the ground state when
$\lambda_3(\lambda_2)$
is varied from $1$ to $0$, and the gap must vanish at a transition point.
In Fig.~\ref{fig:lambda_gap} we also plot the gap as the function of $\lambda_{2,3,}$
when such a wrong superlattice is imposed. We observe that the gap does vanish before
$\lambda_{2,3}$ reaches zero, which provides an alternative confirmation for
the nature of the phases.

\begin{figure}[tbp]
\begin{center}
\includegraphics[width=0.9\columnwidth]{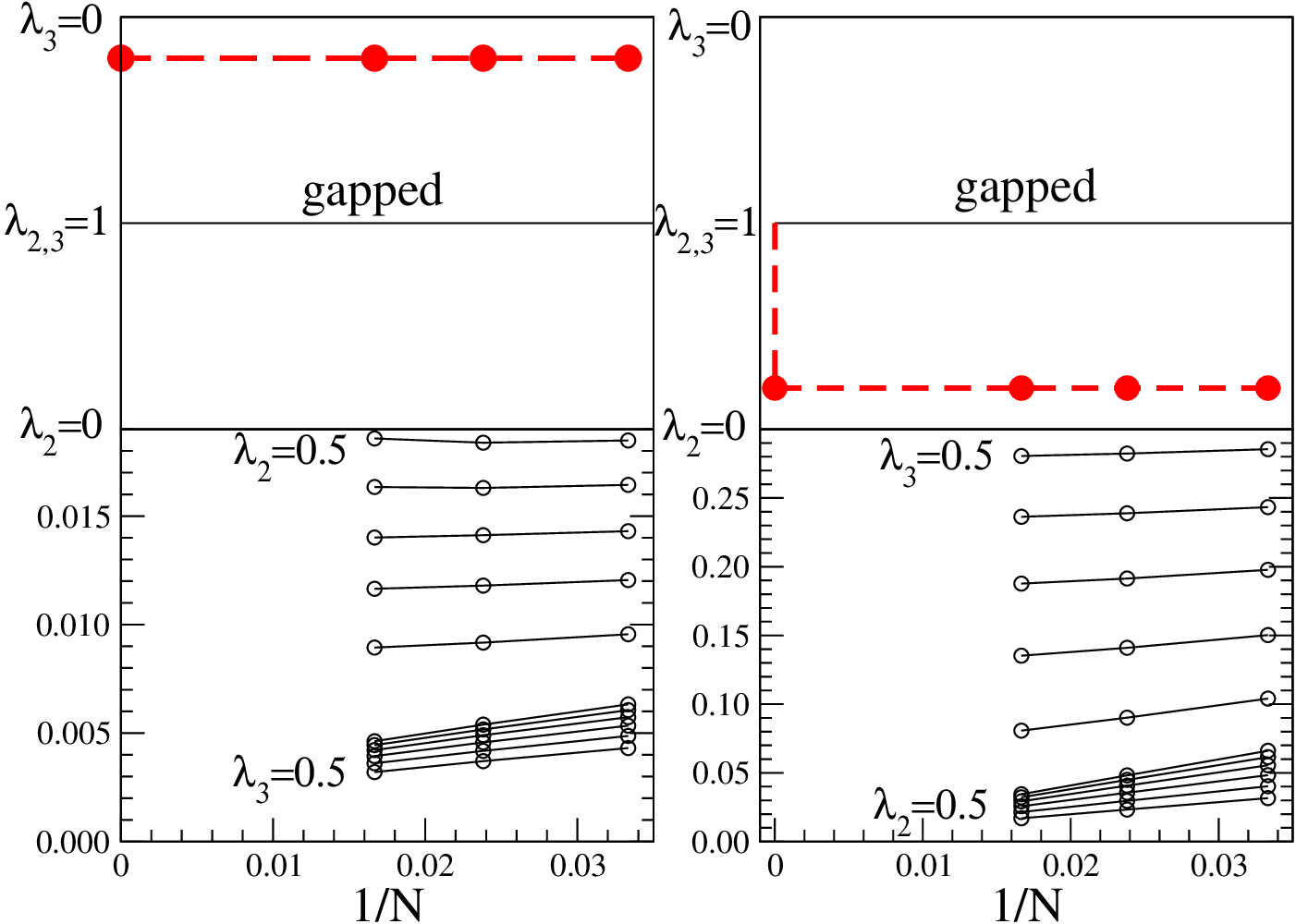}
\caption{(Color online)  ``Phase-diagram" (top panels) and
energy gaps (lower panels) as a function of $N$ and $\lambda_{2,3}$.
The red dotted lines indicate where the system becomes gapless.
Left column, parameters according to $^{23}$Na; right column,
the point $(0,1,0)$.
}
\label{fig:scaling}
\end{center}
\end{figure}

 Now we consider $N \to \infty$.  For the dimerized phase, our description
 above remains valid, except that the magnitude of the gap
 becomes smaller.
 (Fig \ref{fig:lambda_gap}, full symbols).  That is, the system remains gapped,
 unless a sufficiently strong ``wrong'' superlattice is imposed.  For the
 trimerized phase, however, the situation is slightly different.  Without
 any superlattice, the gap vanishes as $1/N$ as $N \to \infty$.
 That is, the system becomes gapless.
 Also, trimer order parameter defined by the appropriate sums and
 differences of the neighboring spin-spin correlation vanishes as
 $N \to \infty$ (not shown).
 \cite{1Dtrimer}.
 Nevertheless, for any finite $\lambda_3$, we find that the gap remains finite
 as $N \to \infty$.  The gapped region as a function of $\lambda_{2,3}$ and
 $1/N$, as well as the gap dependence, are shown in Fig \ref{fig:scaling}.

It is natural to ask what are the expected phases of some of the available spin-2 elements.
Using scattering lengths
available in the literature \cite{as},
we find that $^{83}$Rb and $^{23}$Na should be in the ferromagnetic and dimerized state
respectively. For $^{87}$Rb, however,
the spin-spin correlation only shows a very weak
trimer-like pattern (not shown).
We hence resort to use
the method of imposing superlattices. We impose both
trimerization and dimerization superlattices and calculate
the gap as a function of $\lambda_{2,3}$.
We find that the gap monotonically increases under trimerization superlattice but decreases
to zero under dimerization superlattice. This strongly suggests that the $^{83}$Rb is
indeed in trimerized phase.


Spin-2 bosons in 1D insulating lattice has been studied theoretically before in Ref.~\cite{Eckert07}.
Qualitatively, their picture agrees with ours for the lower left and lower right corners
of the triangle, where they find ferromagnetic and dimerized phase respectively.  However,
our results are qualitatively different for the rest of the phase diagram.  Near the
center of the triangle but excluding the ferromagnetic regime,
they claim to have nematic and cyclic phases. We do not find these states to be stable.
Our result is also different in the region in the upper corner.  There, they proposed
in fact two phases for this region, which they named ``cyclic'' para-dimers and
``para-dimerized'' para-dimers.  In their picture, every two sites form an effective
spin 2 ``para-dimers'', and these para-dimers either form a cyclic state, or
more complicated composite objects which eventually have a cyclic order parameter.
Trimerization was not discussed in their paper.


Lastly we discuss how our picture of the dimerized and trimerized state
can be tested experimentally.
There have now been many discussions on how to detect spin ordering for atomic gases
in optical lattices \cite{Brennen07,optics}.  We here, however, would like to point
out a simple scheme particularly suitable for our states based on the adiabatic
deformation of Hamiltonian by imposing superlattices as just discussed.
Superlattices can easily be created \cite{Peil03},
and actually has already been utilized for preparation and detection of
certain spin-ordering states \cite{Trotzky10}.
In our scheme, by applying an optical potential with two or three lattice spacing and
changing the laser intensity for this optical potential,
one can tune the parameter $\lambda_{2,3}$ defined above since the tunneling
amplitude $t$ decreases with increasing potential barrier.
A trimerized(dimerized) state is then characterized by a monotonic increase of the gap
as a function of $\lambda_3(\lambda_2)$ and the gap closing before $\lambda_2(\lambda_3)$ reaches zero.
Furthermore, by reversing the above argument, a
trimerization(dimerization) superlattice
can also allow us to prepare the trimerized(dimerized) state by first preparing the
corresponding system with $\lambda_{3,2}=0$ and then gradually increasing it until it
reaches $1$.


In summary we have studied the phase diagram of spin-2 bosons in 1D optical lattice
using non-Abelian DMRG. We identify three possible phases, namely ferromagnetic,
dimerized, and trimerized, where the trimerized phase was not proposed in the literature.
We demonstrate that by imposing proper superlattices and
observing the excitation gap, the phase of interest can be identified or prepared
based on the adiabatic connection of the ground state. Such a procedure can be
implemented in cold-atom experiments and provide a simple scheme to test the
phase diagram experimentally.

We acknowledge inspiring conversations with M. A. Cazalilla.
This research was supported by the  National Science Council of Taiwan.



\begin{thebibliography}{plain}



\bibitem{Bloch08}
I. Bloch, J. Dalibard and W. Zwerger, Rev. Mod. Phys. {\bf 80}, 885 (2008).

\bibitem{BEC}
M. H. Anderson {\it et al}, Science {\bf 269}, 98 (1995);
C. C. Bradley {\it et al} Phys. Rev. Lett., {\bf 75}, 1687 (1995);
K. B. Davis {\it et al} {\it ibid} {\bf 75}, 3969 (1995).

\bibitem{Greiner02}
M. Greiner {\it et al}, Nature, {\bf 415}, 39 (2002).

\bibitem{S1}
J. Stenger {\it et al}, Nature (London), {\bf 396}, 345 (1998);
M.-S. Chang {\it et al}, Phys. Rev. Lett. {\bf 92}, 140403 (2004);
M.-S. Chang {\it et al}, Nature Phys. {\bf 1}, 111 (2005).

\bibitem{SH}
S.-K. Yip, Phys. Rev. Lett. {\bf 90}, 250402 (2003);
A. Imambekov, M. Lukin and E. Demler, Phys. Rev. A {\bf 68}, 063602 (2003).

\bibitem{XXZ}
A. B. Kuklov and B. V. Svistunov, Phys. Rev. Lett. {\bf 90}, 100401 (2003);
E. Altman {\it et al}, New J. Phys. {\bf 5}, 113 (2003).

\bibitem{Schmaljohann04}
H. Schmaljohann {\it et al}  Phys. Rev. Lett. {\bf 92}, 040402 (2004).

\bibitem{Kuwamoto04}
T. Kuwamoto {\it et al}, Phys. Rev. A {\bf 69}, 063604 (2004).

\bibitem{Widera06}
A. Widera {\it et al}, New J. Phys. {\bf 8}, 152 (2006).

\bibitem{Tojo09}
S. Tojo {\it et al}, Phys. Rev. A {\bf 80}, 042704 (2009).

\bibitem{Ciobanu00}
C. V. Ciobanu, S.-K. Yip, and T.-L. Ho, Phys. Rev. A {\bf 61}, 033607 (2000).

\bibitem{Koashi00}
M. Koashi and M. Ueda, Phys. Rev. Lett. {\bf 84}, 1066 (2000)

\bibitem{Zhou06}
F. Zhou and G.W. Semenoff, Phys. Rev. Lett. {\bf 97}, 180411 (2006);

\bibitem{Barnett06}
R. Barnett, A. Turner, and E. Demler, Phys. Rev. Lett. {\bf 97}, 180412 (2006).

\bibitem{notes}
The polar phase is also called the nematic phase \cite{Zhou06,Barnett06},
whereas the cyclic phase is also named the tetrahedral state due to their symmetries \cite{Barnett06,Yip07}.

\bibitem{Yip07}
S.-K. Yip, Phys. Rev. A {\bf 75}, 023625 (2007)

\bibitem{fluct}
J. L. Song {\it et al}, Phys. Rev. Lett. {\bf 98}, 160408 (2007);
A. M. Turner {\it et al}, {\it ibid}, 190404 (2007).


\bibitem{noteH}
There are a number of recent papers on 1D spin-2 Hamiltonian,
e.g. \cite{Zang10}, but their parameter regimes are rather different
from the one discussed here, {\it i.e.} $\epsilon_0, \epsilon_2, \epsilon_4 < 0$.


\bibitem{Zang10}
J. Zang {\it et al}, Phys. Rev. B {\bf 81}, 224430 (2010);
H.-C. Jiang, {\it ibid}, {\bf 82}, 220403 (2010).

\bibitem{quad}
We assumed that the magnetic field is sufficiently well-screened
that the quadratic Zeeman effect can be ignored.
For discussions  of its effect in higher dimensional lattice,
see e.g. M.-C. Chung and S.-K. Yip, Phys. Rev. A {\bf 80}, 053615 (2009);
M. Snoek {\it et al}, {\it ibid}, {\bf 80}, 053618 (2009).


\bibitem{DMRG}
I. P. McCulloch and M. Gulacsi, Europhys. Lett. {\bf 57}, 852 (2002)
 I. P. McCulloch, J. Stat. Mech.: Theor. Exp. P10014 (2007)

\bibitem{1Ddimer}
M. Rizzi, D. Rossini, G. De Chiara, S. Montangero and R. Fazio, Phys. Rev. Lett. {\bf 95}, 240404 (2005)
K. Harada, N. Kawashima and M. Troyer, J. Phys. Soc. Jpn., {\bf 76}, 013703 (2007); and references therein.

\bibitem{1Dtrimer}
Our trimerized state thus has some strong similarities to the trimerized state found for the spin-1 chain
(G. F\'{a}th and S\'{o}lyom, Phys. Rev. B {\bf 47}, 872 (1993);
C. Itoi and M.-H. Kato, {\it ibid}, {\bf 55}, 8295 (1997);
A. L\"{a}uchli {\it et al}, {\it ibid}, {\bf 74}, 144426 (2006)).
We note also that in the parameter range where this trimerized phase is stable,
three spin-1 particles also form a unique spin singlet.
We remark however here that the parameters needed to realize this spin-1 trimerized phase
is quite opposite to those which can be obtained for spin-1 Bosons
\cite{SH}.

\bibitem{as}
For $^{83}$Rb, $^{87}$Rb and $^{23}$Na, we used the scattering lengths in
\cite{Ciobanu00} provided originally by J. Burke and C. Greene.

\bibitem{Eckert07}
K. Eckert {\it et al}, New J. Phys. {\bf 9}, 133 (2007).

\bibitem{Brennen07}
G. K. Brennen, A. Micheli and P. Zoller, New J. Phys. {\bf 9}, 138 (2007).

\bibitem{optics}
K. Eckert {\it et al}, Phys. Rev. Lett. {\bf 98}, 100404 (2007);
I. de Vega {\it et al}, Phys. Rev. A {\bf 77}, 051804(R) (2008).


\bibitem{Peil03}
S. Peil {\it et al}, Phys. Rev. A {\bf 67}, 051603 (2003).

\bibitem{Trotzky10}
S. Trotzky {\it et al}, Phys. Rev. Lett. {\bf 105}, 265303 (2010).







\end{thebibliography}
\end{document}